\documentclass[english,prl,superscriptaddress,twocolumn]{revtex4}
\usepackage[T1]{fontenc}
\usepackage[utf8]{inputenc}
\usepackage{babel}
\usepackage{units}
\usepackage{graphicx}
\usepackage{esint}
\usepackage[unicode=true,pdfusetitle,
 bookmarks=true,bookmarksnumbered=false,bookmarksopen=false,
 breaklinks=false,pdfborder={0 0 1},backref=false,colorlinks=false]
 {hyperref}

\makeatletter
\@ifundefined{textcolor}{}
{%
 \definecolor{BLACK}{gray}{0}
 \definecolor{WHITE}{gray}{1}
 \definecolor{RED}{rgb}{1,0,0}
 \definecolor{GREEN}{rgb}{0,1,0}
 \definecolor{BLUE}{rgb}{0,0,1}
 \definecolor{CYAN}{cmyk}{1,0,0,0}
 \definecolor{MAGENTA}{cmyk}{0,1,0,0}
 \definecolor{YELLOW}{cmyk}{0,0,1,0}
}

\usepackage{babel}

\makeatother

\begin{document}

\title{Giant Phonon Anomaly associated with Superconducting Fluctuations
in the Pseudogap Phase of Cuprates}

\author{Y. H. Liu}

\affiliation{Institut für Theoretische Physik, ETH Zürich, CH-8093, Zürich, Switzerland}

\author{R. M. Konik}

\affiliation{Condensed Matter Physics and Material Science Department, Brookhaven
National Laboratory, Upton, NY 11973}

\author{T. M. Rice}

\affiliation{Institut für Theoretische Physik, ETH Zürich, CH-8093, Zürich, Switzerland}

\affiliation{Condensed Matter Physics and Material Science Department, Brookhaven
National Laboratory, Upton, NY 11973}

\author{F. C. Zhang}

\affiliation{Department of Physics, Zhejiang University, Hangzhou, China}

\affiliation{Collaborative Innovation Center of Advanced Microstructures, Nanjing,
China}

\maketitle
\textbf{The opening of the pseudogap in underdoped cuprates breaks
up the Fermi surface, which may lead to a breakup of the $d$-wave
order parameter into two subband amplitudes and a low energy Leggett
mode due to phase fluctuations between them. This causes a large increase
in the temperature range of superconducting fluctuations with an overdamped
Leggett mode. Almost resonant scattering of inter-subband phonons
to a state with a pair of Leggett modes causes anomalously strong
damping. In the ordered state, the Leggett mode develops a finite
energy, suppressing the anomalous phonon damping but leading to an
anomaly in the phonon dispersion. }

The unexpected discovery of a giant anomaly in the dispersion of low
energy phonons (GPA) in underdoped pseudogap cuprates has stimulated
reconsideration of the role of phonons in cuprate high-$T_{c}$ superconductors~\cite{Chang2012,Ghiringhelli2012,Achkar2012,LeTacon2014,Blackburn2013}.
Recently many groups have proposed these anomalies are caused by other
electronic instabilities e.g. charge density wave (CDW) order and
also pair density wave (PDW) order, which compete with the uniform
$d$-wave pairing state~\cite{Hayward2014,Efetov2013,Bulut2013,Melikyan2014,Fradkin2014,Tsvelik2014,Chowdhury2015,Wang2015}.
A novel proposal has been put forward by Lee, who argues that Amperean
pairing occurs in the pseudogap phase leading to an instability towards
PDW and also CDW order~\cite{Lee2014}.

Two recent studies of YBCO samples covering a range of hole densities,
found an onset hole density $p_{c1}\sim0.18$ for the lattice anomalies,
which coincides with the onset of the pseudogap~\cite{Blanco-Canosa2014,Wu2015}.
Early photoemission (ARPES) experiments found that the onset of the
pseudogap is characterized by a breakup of the Fermi surface into
4 pieces centered on the nodal directions~\cite{Norman1998}. A rapid
expansion of the temperature range of superconducting fluctuations
above the transition temperature for long range superconductivity,
$T_{c}\left(p\right)$, is also observed~\cite{Dubroka2011}. This
combination of the onset in hole doping, coinciding with Fermi surface
breakup, and the onset in temperature, coinciding with the onset of
SC fluctuations, leads us to examine possible consequences of the
special disconnected nature of the Fermi surface in the pseudogap
phase, on $d$-wave superconductivity. We find that superconducting
fluctuations in an extended temperature above $T_{c}$ can result
as a special feature of $d$-wave superconductivity in the presence
of the pseudogap. We shall show below that these fluctuations in turn
can couple to finite wavevector phonons leading to GPA. In our model
the lattice fluctuations are dynamic as argued by LeTacon et al~\cite{Blanco-Canosa2014,LeTacon2014}
but random static CDW can still be induced by local perturbations,
such as the random acceptors in nearly all underdoped cuprates~\cite{Wu2015}.
The stoichiometric underdoped cuprate $\mbox{Y}\mbox{Ba}_{2}\mbox{Cu}_{4}\mbox{O}_{8}$
is an exception. The NMR/NQR experiments by Suter et al~\cite{Suter2000}
found dynamic charge fluctuations but no static lattice ordered modulation
in agreement with earlier NMR experiments~\cite{Machi1991,Mangelschots1992}.
A recent detailed NMR study of YBCO found evidence for static lattice
distortions, possibly induced around lattice imperfections by GPA.
They also did not report systematic splitting of the NMR lines which
would be evidence for long range ordered CDW~\cite{Wu2015}. 

In our microscopic model the key ingredient is the novel breakup of
the full Fermi surface into 4 disconnected pieces that characterizes
the pseudogap phase. In the superconducting phase, the 4 pieces combine
to give 2 sets of Cooper pairs with nodes along $\left(1,1\right)$
and $\left(1,\bar{1}\right)$. This breakup opens the possibility
of superconducting phase fluctuations not just of the overall Josephson
phase, but between the disconnected sets of Cooper pairs. The possibility
of phase fluctuations between separated pieces of the Fermi surface
in multiband $s$-wave superconductors, i.e. the Leggett mode (LM),
was studied by Leggett~\cite{Leggett1966} many years ago. The superconductivity
in cuprates has $d$-wave symmetry with nodes on the pockets and the
accompanying sign changes reduce the net coupling between the two
sets. We propose this can lead to a low energy LM.

\section*{Representation of Fermi surface}

\begin{figure}
\includegraphics[width=8.5cm]{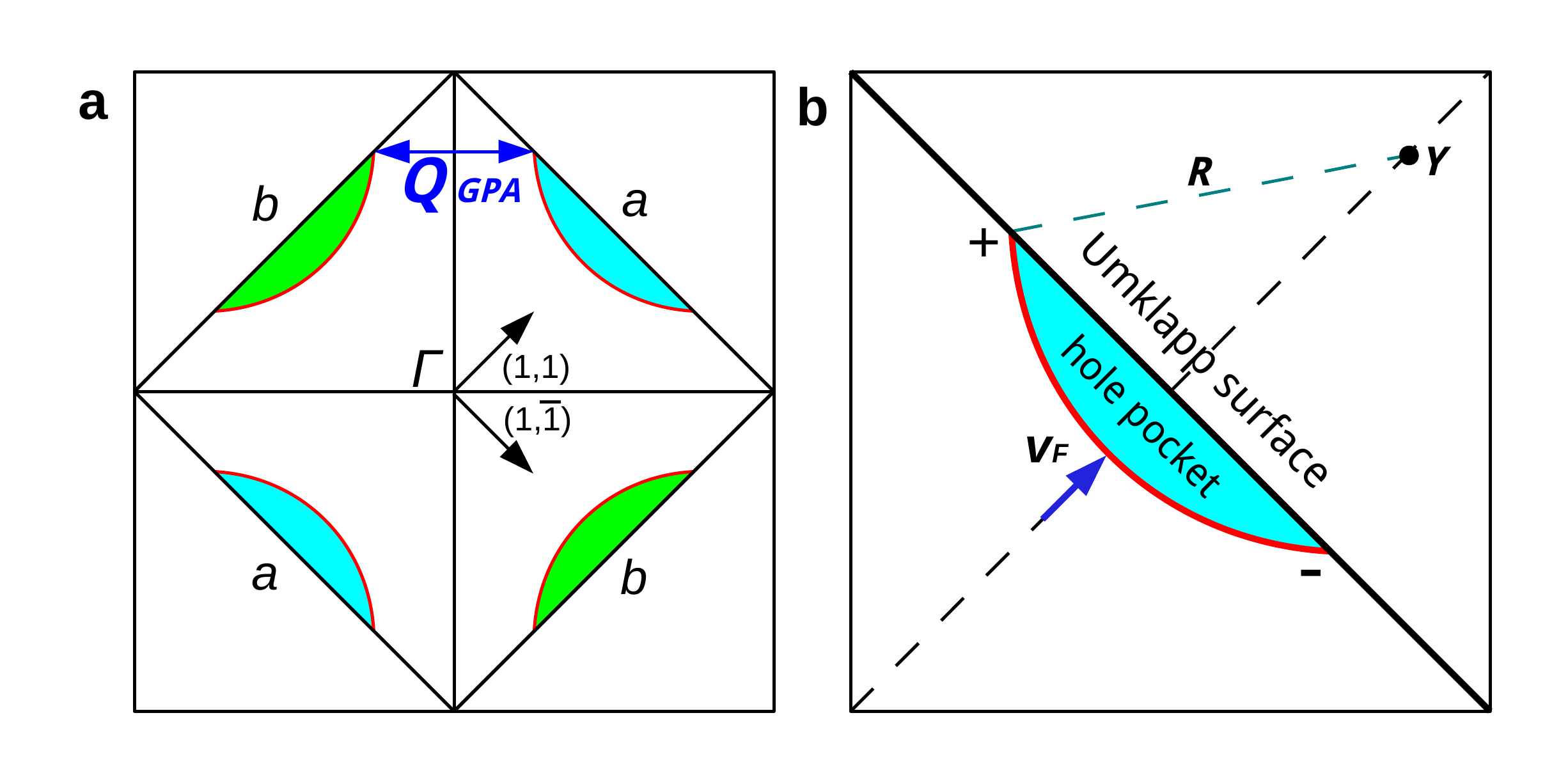}\caption{\textbf{Representation of the band structure by Fermi arcs.} \textbf{a,}
The breakup of Fermi surface to 2 subbands $a$ \& $b$ in $\left(1,1\right)$
\& $\left(1,\bar{1}\right)$ directions. $\mathbf{Q}_{\mathrm{GPA}}$
is a wavevector connecting the two subbands, which is also the wavevector
of phonon anomaly. \textbf{b,} Simplified model of Fermi arcs. Each
Fermi arc is represented by a circular arc (shown red) with center
$\mathbf{Y}$, radius $R$, and terminates at the Umklapp surface.
The Fermi velocity $\mathbf{v}_{F}$ (blue arrow) is assumed to have
constant magnitude on the whole arc. $\mathbf{Y}$ is uniquely determined
by the choices of the wavevector between arc tips to be $0.51\pi$,
and the hole concentration $p=11.5\%$. These are typical values in
ARPES experiment. $\pm$ denotes the sign of symmetry factor $\gamma_{\mathbf{p}}$.
\label{fig:band}}
\end{figure}

One-loop Renormalization Group calculations of the single band Hubbard
model near $\nicefrac{1}{2}$-filling, show that at low energies the
Coulomb repulsion develops substantial structure in $\mathbf{k}$-space,
peaking at momentum transfers $\left(\pi,\pi\right)$ in both the
particle-hole spin triplet and particle-particle singlet channels~\cite{Honerkamp2001,Honerkamp2002}.
The latter drives $d$-wave pairing with maximum amplitudes at antinodal.
Yang et al~\cite{Yang2006,Rice2012} argued the transition into the
pseudogap phase is driven by increasing Umklapp scattering, which
converts the antinodal gaps to insulating. They put forward an ansatz
based on a modified BCS self energy with an energy gap pinned to the
Umklapp surface leading to anisotropic Fermi pockets centered on the
nodal directions. Later detailed ARPES experiments confirmed anisotropic
nodal Fermi pockets~\cite{Yang2011}. The superconducting complex
pairing amplitude is confined to 2 disconnected pairs of pockets centered
on the $\left(1,1\right)$ \& $\left(1,\bar{1}\right)$ directions,
illustrated in Fig.~\ref{fig:band}. We shall refer to these as subband
$a$ \& $b$ respectively. An examination of the pair scattering processes
shows that there are intra-subband $\left(\pi,\pi\right)$ processes.
But the inter-subband processes connecting subbands $a$ \& $b$ involve
cancellations between pairing and depairing scattering processes with
similar wavevectors in a $d$-wave state. For this reason when we
write the pairing scattering in terms of intra-subband and inter-subband
processes for a $d$-wave state, we propose that the intra-subband
pairing is stronger than the inter-subband pairing, similar to the
pairing model examined by Leggett~\cite{Leggett1966}.

For simplicity the anisotropic nodal Fermi pockets will be represented
as Fermi arcs (see Fig.~\ref{fig:band}) with a constant Fermi velocity
along the arc. The parameters are chosen from a recent paper by Comin
et al~\cite{Comin2014}.

\section*{Leggett Mode and Fluctuations}

Assuming that both intra-subband and inter-subband couplings $U$
and $J$ are separable $d$-wave forms with symmetry factor, $\gamma_{\mathbf{p}}$,
allows us to obtain the fluctuation pair propagator from the following
Bethe-Salpeter equation 
\begin{equation}
L_{q}=\left(\begin{array}{cc}
U & J\\
J & U
\end{array}\right)-\left(\begin{array}{cc}
U & J\\
J & U
\end{array}\right)\left(\begin{array}{cc}
\pi_{q}^{\left(a\right)} & 0\\
0 & \pi_{q}^{\left(b\right)}
\end{array}\right)L_{q}\label{eq:BSE}
\end{equation}
for $L_{q}$, in terms of which the full anisotropic fluctuation is
written as $L_{pp'q}=\gamma_{\mathbf{p}}L_{q}\gamma_{\mathbf{p}'}$
with $q=\left(\mathbf{q},iq_{0}\right)$. The electronic bubble $\pi_{q}^{\left(i\right)}=\frac{1}{\beta V}\sum_{\mathbf{p},i\omega}^{\left(i\right)}\gamma_{\mathbf{p}}^{2}G{}_{\frac{\mathbf{q}}{2}+\mathbf{p},iq_{0}+i\omega}G_{\frac{\mathbf{q}}{2}-\mathbf{p},-i\omega}$
is defined on both sets of Fermi arcs for $i=a,b$. We denote zero-temperature,
finite-temperature, and retarded Green's functions by $G_{\mathbf{p},\omega}$,
$G_{\mathbf{p},i\omega}$, and $G_{\mathbf{p},\omega}^{R}$ respectively,
similarly for other quantities. In the temperature region $T_{c}<T<T_{o}$
(onset of superconducting fluctuations), the electron's retarded Green's
function takes the form $G_{\mathbf{p},\omega}^{R}=\left[\omega-\epsilon_{\mathbf{p}}+i\Gamma^{\left(e\right)}\right]^{-1}$,
where $\epsilon_{\mathbf{p}}$ is the Fermi arc dispersion and $\Gamma^{\left(e\right)}=aT+bT^{2}$
is a temperature dependent quasiparticle damping~\cite{Buhmann2013}.
The retarded propagator for the fluctuating LM is derived to be 
\begin{equation}
L_{\mathbf{q},\omega}^{R}=\frac{T}{cN_{0}}\frac{1}{i\omega-\Gamma_{\mathbf{q}}^{\left(\mathrm{LM}\right)}}\left(\frac{1}{2}-\frac{1}{2}\sigma_{x}\right).
\end{equation}
See the supplementary information (SI) for details. This form of the
fluctuation propagator describes an overdamped bosonic mode. Here
$N_{0}$ is the density of states per spin at the Fermi energy for
one pair of arcs, the constant $c=\frac{1}{4\pi}\psi'\left[\frac{1}{2}+\frac{1}{2\pi}\left(a+bT_{c}\right)\right]$
with $\psi\left(x\right)$ the digamma function. The damping of the
LM $\Gamma_{\mathbf{q}}^{\left(\mathrm{LM}\right)}=\tau^{-1}+Dq^{2}$,
in which the inverse relaxation time $\tau^{-1}=\frac{T}{c}\log\frac{T}{T_{c}}+2bT\left(T-T_{c}\right)+\frac{T}{cN_{0}}2\left|J\right|/\left(U^{2}-J^{2}\right)$
and the diffusion constant $D=-\frac{v_{F}^{2}}{16\pi T}\psi''\left[\frac{1}{2}+\frac{1}{2\pi}\left(a+bT\right)\right]/\psi'\left[\frac{1}{2}+\frac{1}{2\pi}\left(a+bT_{c}\right)\right]$
are functions of the temperature. The uniform damping rate $\tau^{-1}$
decreases as long range order at $T_{c}$ is approached.

We take zero temperature as a representative case for the ordered
phase. In this case the LM is derived similarly to equation~(\ref{eq:BSE}),
but with a different form of the electronic bubble~\cite{Leggett1965,Leggett1966a,Leggett1966}
$\pi_{q}^{\left(i\right)}=\frac{1}{V}\sum_{\mathbf{p}}^{\left(i\right)}\int\frac{d\omega}{2\pi i}\gamma_{\mathbf{p}}^{2}\left(G_{\frac{\mathbf{q}}{2}+\mathbf{p},\frac{q_{0}}{2}+\omega}G_{\frac{\mathbf{q}}{2}-\mathbf{p},\frac{q_{0}}{2}-\omega}\right.+\left.F_{\frac{\mathbf{q}}{2}+\mathbf{p},\frac{q_{0}}{2}+\omega}F_{\frac{\mathbf{q}}{2}-\mathbf{p},\frac{q_{0}}{2}-\omega}\right)$.
$G_{\mathbf{p},\omega}=\left(\omega+\epsilon_{\mathbf{p}}\right)/Z$
and $F_{\mathbf{p},\omega}=\Delta_{\mathbf{p}}/Z$ are the normal
and anomalous Green's functions, where $Z=\omega^{2}-E_{\mathbf{p}}^{2}+i\delta$
and $\delta\rightarrow0^{+}$. $E_{\mathbf{p}}=\sqrt{\epsilon_{\mathbf{p}}^{2}+\Delta_{\mathbf{p}}^{2}}$
is the quasiparticle energy and $\Delta_{\mathbf{p}}=\gamma_{\mathbf{p}}\Delta$
is the $d$-wave gap function. It follows (see SI) \textbf{
\begin{equation}
L_{\mathbf{q},\omega}=\frac{4\Delta^{2}}{N_{0}}\frac{1}{\omega^{2}-\omega_{\mathbf{q}}^{2}+i\delta}\left(\frac{1}{2}-\frac{1}{2}\sigma_{x}\right)
\end{equation}
}with a LM dispersion $\omega_{\mathbf{q}}^{2}=\omega_{0}^{2}+\frac{1}{2}v_{F}^{2}q^{2}$.
In this case, the LM is a coherent bosonic mode with infinite lifetime.
The LM is gapped, and its frequency at zero momentum satisfies $\omega_{0}^{2}=\frac{4\Delta^{2}}{N_{0}}2\left|J\right|/\left(U^{2}-J^{2}\right)$.
Usually the ratio $\frac{\omega_{0}}{2\Delta}<1$.

\section*{Phonon Self Energy}

\begin{figure}
\includegraphics[width=9cm]{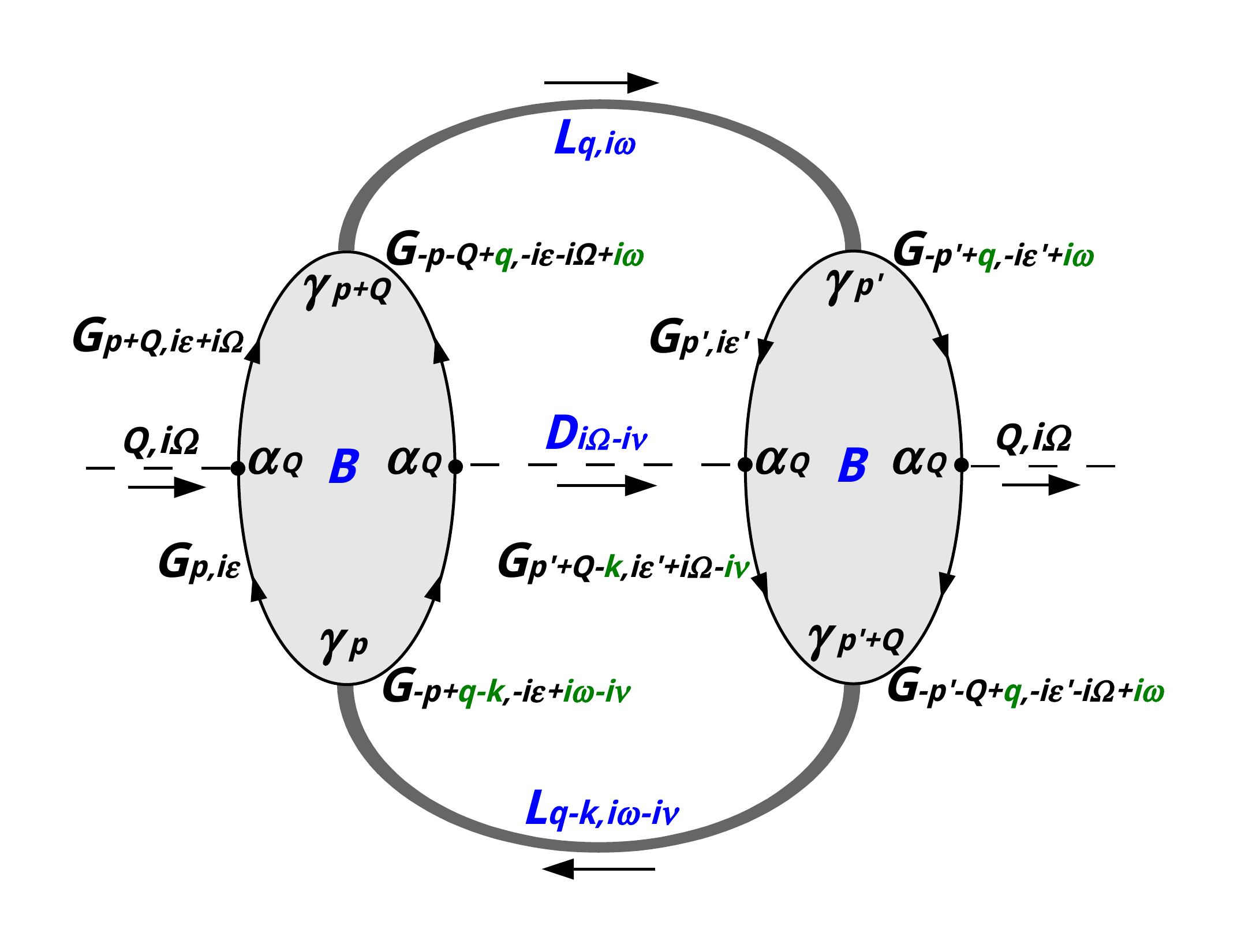}\caption{\textbf{Feynman diagram of phonon self energy due to the interaction
with Leggett mode.} $B$ is the effective interaction vertex consisting
of a particle-particle electronic bubble ($G$ is the electron Green's
function). The intermediate state consists of two Leggett modes ($L$)
and one phonon ($D$). Due to the separation of energy scales between
Leggett mode and the electron, all quantities in green are neglected.
\label{fig:diagram}}
\end{figure}

The $\mathbf{k}$-space separation of the two bands does not move
the LM away from $\mathbf{q}=0$, since this phase mode involves the
transfer of zero momentum Cooper pairs between the subbands. Nonetheless
it involves moving charges between the subbands. Absorption and emission
of phonons with the appropriate wavevectors also causes a charge transfer,
but now as single quasiparticles, between the two subbands. Therefore
it is not unexpected that a coupling between these processes should
exist. In particular, we find the coupling is largest in the temperature
region $T\sim T_{c}$, where the LM drops to zero energy and becomes
overdamped. To this end we consider the process outlined in Fig.~\ref{fig:diagram},
where an incoming phonon is scattered to a nearby phonon wavevector
with emission and absorption of LM fluctuations. Such a process does
not occur in standard superconductors but can exist here because of
a soft overdamped LM for $T_{c}<T<T_{o}$. Below we summarize calculations
of the phonon self energy in two temperature regions.

\begin{figure}
\includegraphics[width=8.5cm]{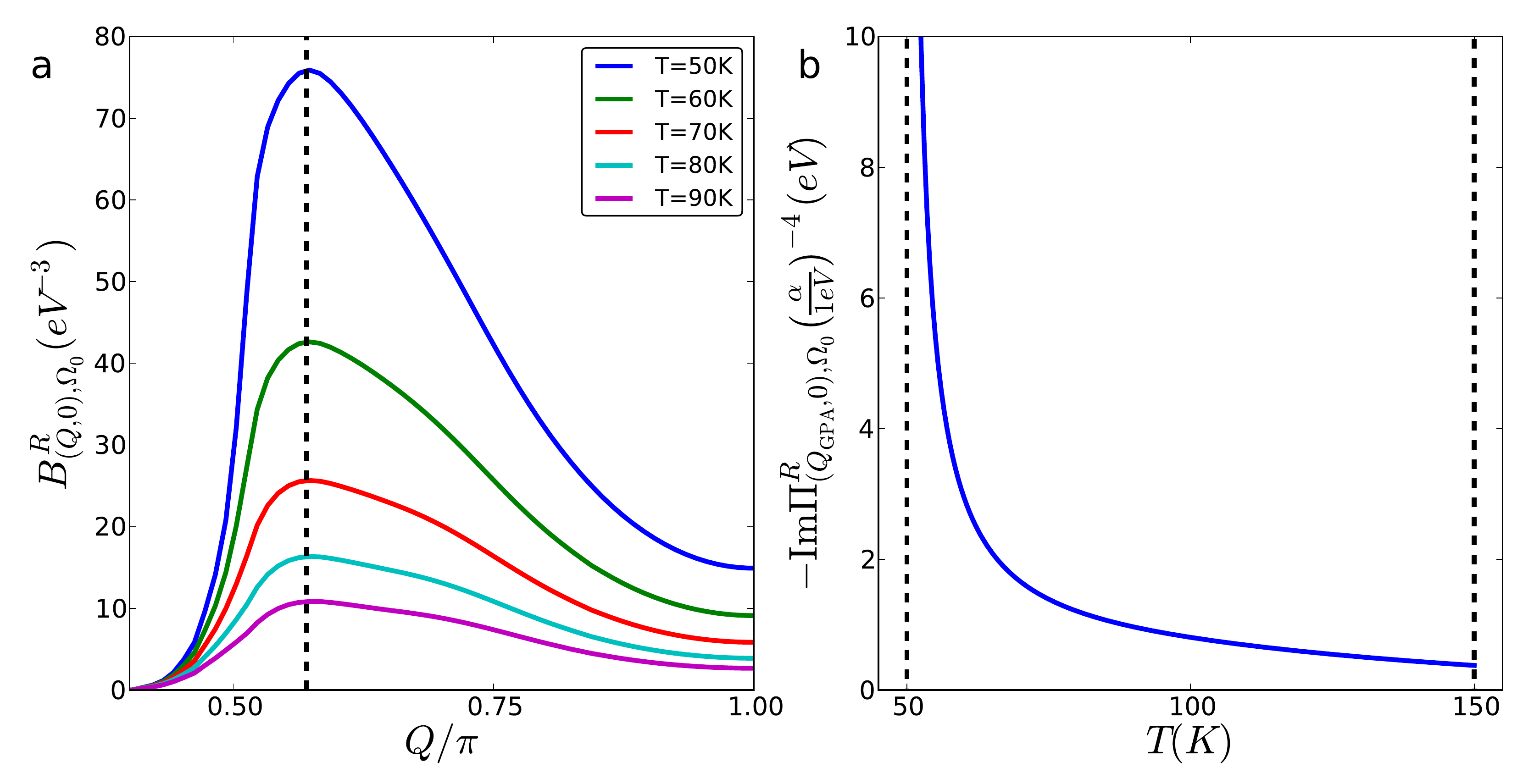}\caption{\textbf{GPA in the fluctuation region.} \textbf{a,} Temperature and
momentum dependence of the effective interaction vertex $B$, with
the dashed line marking $Q_{\mathrm{GPA}}$.\textbf{ b,} The anomalous
phonon damping. Parameters: Fermi velocity $v_{F}=500\mbox{meV}$,
bare phonon frequency $\Omega_{0}=10\mbox{meV}$, ratio between Leggett
mode frequency at $\mathbf{q}=0$ and the superconducting gap $\frac{\omega_{0}}{2\Delta}=0.1$,
quasiparticle damping $\Gamma^{\left(e\right)}=0.5T+\left(0.3\mbox{meV}^{-1}\right)T^{2}$,
and the long range order temperature $T_{c}=50\mbox{K}$. We used
an energy cutoff of $\pm100\mbox{meV}$ around the Fermi surface,
which does not affect the qualitative feature of GPA. \label{fig:numerics}}
\end{figure}

First we look at the phonon damping in the range of strong SC fluctuations
starting at the onset temperature $T_{o}$ of the SC fluctuations
down to the superconducting transition temperature $T_{c}$. The expression
for the phonon self energy $\Pi$, corresponding to the Feynman diagram
Fig.~\ref{fig:diagram}, follows 
\begin{eqnarray}
\Pi_{\mathbf{Q},i\Omega} & = & 4\alpha_{\mathbf{Q}}^{4}B_{\mathbf{Q},i\Omega}^{2}I_{i\Omega},\quad I_{i\Omega}=\frac{1}{V^{2}}\sum_{\mathbf{q},\mathbf{k}}\mbox{Tr}I_{\mathbf{q},\mathbf{k},i\Omega},\nonumber \\
I_{\mathbf{q},\mathbf{k},i\Omega} & = & \frac{1}{\beta^{2}}\sum_{i\omega,i\nu}L_{\mathbf{q},i\omega}L_{\mathbf{q}-\mathbf{k},i\omega-i\nu}D_{i\Omega-i\nu},\nonumber \\
B_{\mathbf{Q},i\Omega} & = & \frac{1}{\beta V}\sum_{\mathbf{p},i\epsilon}\gamma_{\mathbf{p}}\gamma_{\mathbf{p}+\mathbf{Q}}\nonumber \\
 &  & \times G_{\mathbf{p},i\epsilon}G_{-\mathbf{p},-i\epsilon}G_{\mathbf{p}+\mathbf{Q},i\epsilon+i\Omega}G_{-\mathbf{p}-\mathbf{Q},-i\epsilon-i\Omega},\nonumber \\
\label{eq:self-energy}
\end{eqnarray}
where $\alpha_{\mathbf{Q}}$ is the electron-phonon coupling constant,
$I_{\mathbf{q},\mathbf{k},i\Omega}$ is the frequency summation over
the intermediate state consisting of two LMs and one phonon, $B_{\mathbf{Q},i\Omega}$
is the effective interaction vertex between phonons and LMs, and $D_{i\Omega}=2\Omega_{0}/\left[\left(i\Omega\right)^{2}-\Omega_{0}^{2}\right]$
is the bare Green's function for phonons with an assumed flat dispersion
$\Omega_{\mathbf{Q}}=\Omega_{0}$. We have chosen the simplest form
of the effective interaction, $B$ and ignore damping due to quasiparticle
excitations, keeping only $\mbox{Re}B_{\mathbf{Q},\Omega}^{R}$, to
concentrate on the novel phonon damping caused by the presence of
a soft LM. Note, $\mbox{Re}B_{\mathbf{Q},\Omega_{0}}^{R}$ has a strong
dependence on the phonon wavevector $\mathbf{Q}$ (see Fig.~\ref{fig:numerics})
and peaks at a wavevector joining the ends of the arcs, because the
symmetry factors and the available phase space for the transition
at this wavevector are both large. We checked that $\mbox{Im}B_{\mathbf{Q},\Omega_{0}}^{R}\ll\mbox{Re}B_{\mathbf{Q},\Omega_{0}}^{R}$
for this set of parameters. $\mbox{Re}B_{\mathbf{Q},\Omega_{0}}^{R}$
also shows a peak at $\mathbf{Q}=0$ which will be discussed later. 

The effective interaction vertex $B$ involves an integration over
the whole Brillouin zone and all frequencies, while the LM $L$ is
only well-defined for small momenta and frequencies. This leads to
a separation of spatial and temporal scales and enables us to ignore
all small wavevectors and frequencies (marked green in Fig.~\ref{fig:diagram})
in calculating $B$. Aided by the similarity with the Aslamazov-Larkin
diagram~\cite{Aslamasov1968,Schmidt1968,Yip1990,Koshelev2005,Larkin2005},
we perform analytical calculation for $I_{\Omega}^{R}$. For $\left|\Omega-\Omega_{0}\right|\ll\Omega_{0}$,
it follows (see SI for details)

\begin{eqnarray}
I_{\Omega}^{R} & = & \frac{2\pi}{D^{2}}\frac{T^{4}}{c^{2}N_{0}^{2}}\int_{0}^{\infty}dx\, p\left(x\right)\frac{1}{\Omega-\Omega_{0}+\frac{2i\left(1+x\right)}{\tau}},\nonumber \\
p\left(x\right) & = & \int_{0}^{2\pi}d\theta\frac{\arctan\left[\sqrt{\frac{\cos^{2}\theta}{\left(1+x\right)^{2}-x^{2}\cos^{2}\theta}}x\right]}{\sqrt{\cos^{2}\theta\left[\left(1+x\right)^{2}-x^{2}\cos^{2}\theta\right]}}.\label{eq:finite-T-I}
\end{eqnarray}
In this region the on-shell values satisfy $\mbox{Re}I_{\Omega_{0}}^{R}=0$
and $\mbox{Im}I_{\Omega_{0}}^{R}<0$. The temperature dependence of
the imaginary part of the retarded phonon self energy $\mbox{Im}\Pi{}^{R}$
i.e. the phonon damping, is plotted in Fig.~\ref{fig:numerics}.
The self energy has a peak in momentum space at $\mathbf{Q}=\left(Q_{\mathrm{GPA}},0\right)$,
near to the tip to tip wavevector between two sets of Fermi arcs.
Because of the factor $\tau$ in equation~(\ref{eq:finite-T-I}),
the temperature dependence shows anomalous behavior at the long range
critical temperature $T_{c}$, in agreement with experiment~\cite{LeTacon2014}. 

Below $T_{c}$, there is a finite restoring force for inter-subband
phase fluctuations and the LM develops a finite energy at $\mathbf{q}=0$,
which raises the energy of the intermediate state in Fig.~\ref{fig:diagram}.
As a consequence the approximate resonant condition between the incoming
phonon and the intermediate state with a scattered phonon and 2 LMs
no longer holds, leading to a suppression of the phonon damping at
low $T$. The GPA changes its form at $T<T_{c}$ with strongly reduced
damping. An anomaly in the phonon dispersion appears due to virtual
coupling to an excited intermediate state. Treating the low temperature
behavior at $T=0$, the factor from the intermediate state becomes
\begin{eqnarray}
\mbox{Tr}I_{\mathbf{q},\mathbf{k},\Omega} & = & \mbox{Tr}\int\frac{d\omega}{2\pi i}\frac{d\nu}{2\pi i}\, L_{\mathbf{q},\omega}L_{\mathbf{q}-\mathbf{k},\omega-\nu}D_{\Omega-\nu}\nonumber \\
 & = & \frac{4\Delta^{4}}{N_{0}^{2}}\frac{1}{\omega_{\mathbf{q}}\omega_{\mathbf{q}-\mathbf{k}}}\frac{2\left(\Omega_{0}+\omega_{\mathbf{q}}+\omega_{\mathbf{q}-\mathbf{k}}\right)}{\Omega^{2}-\left(\Omega_{0}+\omega_{\mathbf{q}}+\omega_{\mathbf{q}-\mathbf{k}}\right)^{2}+i\delta}\nonumber \\
\end{eqnarray}
where $D_{\Omega}=2\Omega_{0}/\left(\Omega^{2}-\Omega_{0}^{2}+i\delta\right)$.
We conclude that in this region, the on-shell values satisfy $\mbox{Re}I_{\Omega_{0}}^{R}<0$
and $\mbox{Im}I_{\Omega_{0}}^{R}=0$, in agreement with the experiment
for $T<T_{c}$~\cite{LeTacon2014}. 

At long wavelengths, $\mathbf{Q}\sim0$, the long range nature of
the Coulomb interaction suppresses the response in metals at low frequencies
to any perturbation coupling to the total electronic density. The
electron-phonon interaction introduced in equation~(\ref{eq:self-energy})
couples equally to both subbands and as a result the associated scattering
processes are suppressed at $\mathbf{Q}\sim0$. 

\section*{Summary and Conclusions}

As we remarked earlier the transition into the pseudogap phase at
the hole density $p_{c1}\sim0.18$ displays three strong anomalies
simultaneously --- an antinodal insulating energy gap leading to a
breakup of the Fermi surface into 4 nodal pockets, a rapid expansion
of the temperature range of superconducting fluctuations, and the
appearance of a giant phonon anomaly in this temperature range. These
phenomena are unique to the underdoped cuprates. Our aim here is to
put forward a microscopic scenario which explains the interrelation
between these phenomena. The low energy Coulomb interaction must be
strong to drive the partial truncation of the Fermi surface, which
is a precursor to a full Mott gap at zero doping~\cite{Yang2006,Rice2012}.
A weakness of our microscopic scenario is the need to assume a form
for this effective Coulomb interaction. Our choice is guided by the
evolution found by Functional Renormalization Group in the overdoped
density region~\cite{Honerkamp2001,Honerkamp2002}. The persistence
of $d$-wave symmetry even as the maximally gapped antinodal regions
transform from a superconducting to an insulating gap, leads to the
conditions for a low energy LM to emerge. As we discuss above,  this
novel assumption enables us to consistently explain all three anomalies.
In particular we can explain the temperature evolution of the GPA
from increasing damping as $T\rightarrow T_{c}$ from above, to a
GPA with vanishing damping but a dispersion anomaly at $T<T_{c}$.
Here we considered only zero magnetic field. The recent analyses of
quantum oscillation experiments at high magnetic fields are consistently
explained by a coherent orbit around all four arcs, is intriguing~\cite{Sebastian2014}.
It raises the question of the evolution of the LM with increasing
magnetic field for future study.\bibliographystyle{naturemag}
\bibliography{GPAwoURL}

\section*{Acknowledgements}

The authors acknowledge M. Sigrist, A. M. Tsvelik, J. Chang, W. Q.
Chen, J. Gukelberger, D. Manske, M. Troyer, L. Wang, S. Z. Zhang,
and Y. Zhou for helpful discussions. Y.H.L. is supported by ERC grant
No.~290464. R.M.K. is supported by the US DOE under contract number
DE-AC02-98 CH 10886. F.C.Z. is partly supported by NSFC grant 11274269
and National Basic Research Program of China (No. 2014CB921203).

\section*{Author contributions}

The calculations were performed by Y.H.L. with assistance from T.M.R.
All authors discussed the results and took part in the preparation
of the manuscript.

\section*{Additional information}

Supplementary information is available in the online version of the
paper. Reprints and permissions information is available online at
www.nature.com/reprints. Correspondence and requests for materials
should be addressed to T.M.R.

\section*{Competing financial interests}

The authors declare no competing financial interests.
\end{document}